# Topological spin excitations observed in a three-dimensional antiferromagnet

Weiliang Yao[1,#], Chenyuan Li[1,#], Lichen Wang[1,#], Shangjie Xue[1], Yang Dan[1,§], Kazuki Iida[2], Kazuya Kamazawa[2], Kangkang Li[3,4], Chen Fang[3,*], Yuan Li[1,5,*]

[1]International Centre for Quantum Materials, School of Physics, Peking University, Beijing 100871, China
[2]Neutron Science and Technology Centre, Comprehensive Research Organization for Science and Society (CROSS), Tokai, Ibaraki 319-1106, Japan
[3]Beiijng National Laboratory for Condensed Matter Physics, and Institute of Physics, Chinese Academy of Sciences, Beijing 100190, China
[4]University of Chinese Academy of Sciences, Beijing 100049, China
[5]Collaborative Innovation Centre of Quantum Matter, Beijing 100871, China
[#]These authors contributed equally to this work.
[§]Present address: Department of Materials Science and Engineering, University of Illinois at Urbana-Champaign, IL 61801, USA

**Band topology, or global wave-function structure that enforces novel properties in the bulk and on the surface of crystalline materials, is currently under intense investigations for both fundamental interest and its technological promises [1-4]. While band crossing of non-trivial topological nature was first studied in three dimensions for electrons [4-10], the underlying physical idea is not restricted to fermionic excitations [11-15]. In fact, experiments have confirmed the possibility to have topological band crossing of electromagnetic waves in artificial structures [16]. Fundamental bosonic excitations in real crystals, however, have not been observed to exhibit the counterpart under ambient pressure and magnetic field, where the difficulty is in part because natural materials cannot be precisely engineered like artificial structures. Here, we use inelastic neutron scattering to reveal the presence of topological spin excitations (magnons) in a three-dimensional antiferromagnet, $Cu_3TeO_6$, which features a unique lattice of magnetic spin-1/2 $Cu^{2+}$ ions [17]. Beyond previous understanding [17,18], we find that the material's spin lattice possesses a variety of exchange interactions, with the interaction between the ninth-nearest neighbours being as strong as that between the nearest neighbours. Although theoretical analysis indicates that the presence of topological magnon band crossing is independent of model details [15], $Cu_3TeO_6$ turns out to be highly favourable for the experimental observation, as its optical magnons are spectrally sharp and intense due**

**to the highly interconnected spin network and the large magnetic cell. The observed magnon band crossing generally has the form of a special type of $Z_2$-topological nodal lines [19] that are yet to be found in fermion systems, rendering magnon systems a fertile ground for exploring novel band topology.**

Magnons are quantized spin-1 collective excitations from an ordered magnetic ground state. Unlike electrons, magnon band crossing does not require a suitable chemical potential to be observed. However, many magnetic materials have primitive cells that contain too few spins to allow for any band crossing at all. It was previously envisioned that topological magnon band crossing in the form of Weyl points, which occur at generic crystal momenta, would only be possible in the restrictive cases of non-centrosymmetric crystal structures [13] or certain types of ferromagnets [14]. Meanwhile, in light of the additional symmetry requirement for stabilizing Dirac-point-like electronic band crossing [7], the magnetic space groups [20] appear considerably more complicated than the crystallographic space groups [21-23]. Recently, some of us proposed that topological magnon band crossing may occur in a large class of antiferromagnets [15], as long as the *PT* (time reversal followed by space inversion) symmetry is present. This unlocks far more materials to be considered than previously thought.

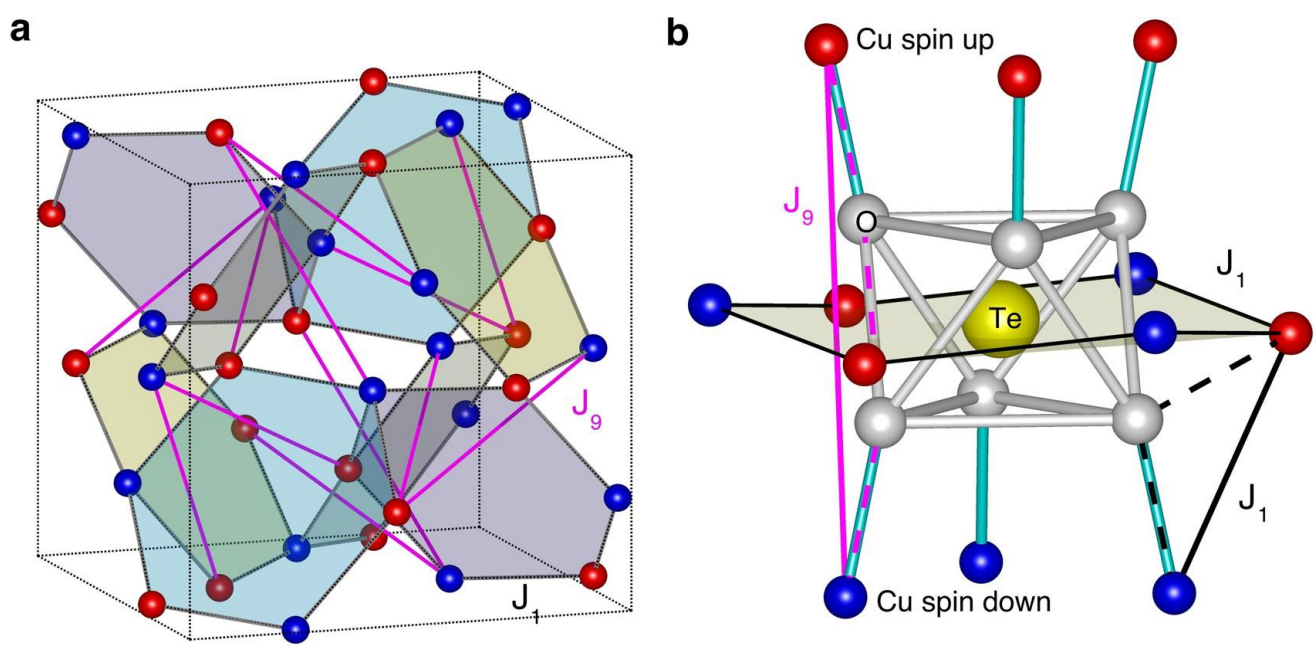

**Figure 1 | Primary magnetic interactions in $Cu_3TeO_6$. a,** The magnetic lattice in a cubic unit cell. Spin-up and spin-down $Cu^{2+}$ are represented in different colours. The nearest-neighbour ($J_1$) and the ninth-nearest-neighbour ($J_9$) interactions constitute a highly interconnected network. **b**, Exchange pathways (dashed lines) of $J_1$ and $J_9$ via oxygen atoms. The relatively straight bond sequence of $J_9$ makes it comparable in strength to $J_1$, despite the much greater distance.

Inelastic neutron scattering (INS) is by far the best method to visualize magnon dispersions in three dimensions. However, experimentally observed magnon bands are not always clear and sharp. Poor sample mosaic, crystal defects, thermal broadening and instrumental resolution all contribute to the experimental linewidth. For antiferromagnets, the intrinsic linewidth (even in perfect crystals at zero temperature) is further increased by quantum fluctuations [24] that are strong in systems with reduced dimensionality, frustrated interactions, and small spin quantum numbers. In fact, quantum fluctuations can be as severe as causing magnons to disintegrate into fractionalized 'spinon' excitations in one [25] and two dimensions [26,27].

The above considerations cast a somewhat pessimistic perspective on the proposal [15] that $Cu_3TeO_6$ is a candidate material for the observation of topological magnons. $Cu_3TeO_6$ develops antiferromagnetic order below $T_N$ = 61 K (Extended Data Fig. 1). The order features a bipartite and predominantly collinear arrangement [17] of spin 1/2 on the $Cu^{2+}$ sub-lattice (Fig. 1a). The apparent difficulty is that each $Cu^{2+}$ has only four nearest neighbours connected by the antiferromagnetic interaction $J_1$, *i.e.*, the coordination number ($N$ = 4) is the same as in a two-dimensional square lattice [18]. Even though a second-nearest-neighbour interaction ($J_2$) might exist, it is expected to be weak, as otherwise the magnetic lattice becomes frustrated [17]. Moreover, spin 1/2 is the extreme case for strong quantum fluctuations.

But as we will show below, the true magnetic interactions in $Cu_3TeO_6$ are dominated by antiferromagnetic $J_9$ (interaction between the ninth-nearest neighbours) and $J_1$ with very similar strengths. Although surprising at first sight, the prominence of $J_9$ can be understood as a super-superexchange interaction [28] that gains its strength from the relatively straight bond sequence Cu-O-O-Cu (Fig. 1b). Additional analyses of the exchange interactions are presented in Extended Data Fig. 2 and Extended Data Table 1. With $J_1 \approx J_9$, the spin lattice is highly interconnected ($N$ = 8) without frustration (Fig. 1a), and quantum fluctuations are strongly suppressed.

Figure 2 presents our measurement of spin excitations in and out of the magnetically ordered state. Well-defined magnon bands are observed at 4 K (Fig. 2a), and they completely collapse into a featureless cloud of paramagnetic excitations at 73 K (Fig. 2b), which is not far above $T_N$. Despite this drastic change, the inelastic spectral weight is approximately conserved (Fig. 2d). Therefore, we can be assured that the transition at $T_N$ is highly three-dimensional (or mean-field-like), and that the spin excitations below $T_N$ are regular magnons that should be describable by a linear spin wave theory (LSWT). The case here is clearly different from some of the best-known low-dimensional spin-1/2 antiferromagnets, where LSWT has only limited success in describing the spin excitations in the ordered state [25-27], and where similar spin excitations are found to persist well into the paramagnetic state [25]. At 4 K, a total of six magnon branches are observed, suggesting that they are all doubly degenerate (since the primitive cell has twelve $Cu^{2+}$ ions) and that the spin Hamiltonian has U(1) spin-rotation symmetry [15]. But this can only be approximately true, because we do observe a small anisotropy gap at the bottom of the 'acoustic' branch (Fig. 2e and f). We will come back to this point later.

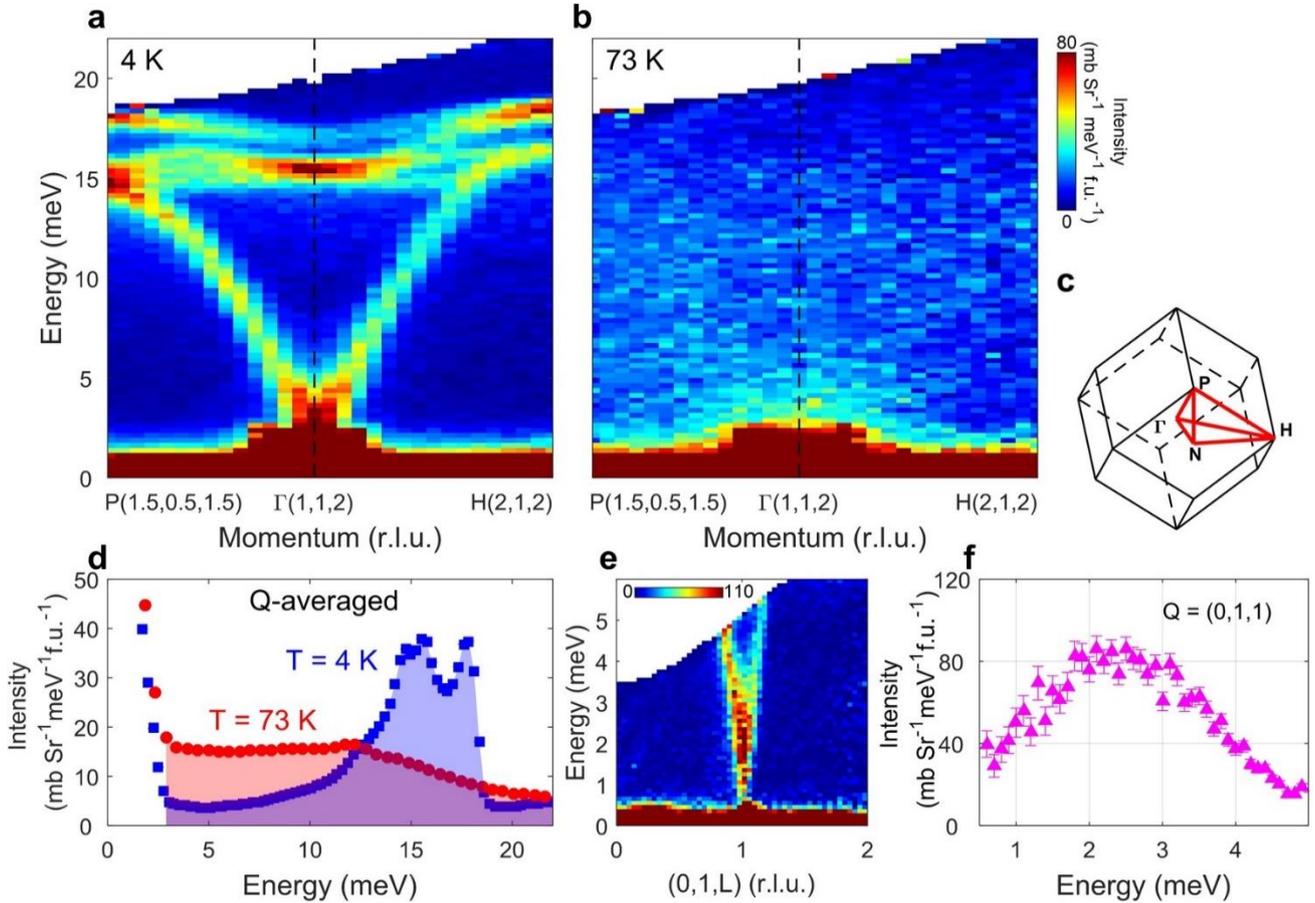

**Figure 2 | Basic properties of spin excitations. a and b,** Representative INS data taken with incident neutron energy $E_i$ = 28 meV in the ordered and the paramagnetic states, respectively. Data are shown near a magnetic wave vector **Q** = (1, 1, 2) (in reciprocal lattice units, r.l.u.). **c,** The Brillouin zone, with high-symmetry lines indicated in red. **d,** Energy distribution of INS intensities averaged over more than ten Brillouin zones. The total spectral weights (shaded areas), mostly magnetic, are the same at both temperatures within 2% accuracy. **e and f,** Data near the bottom of the acoustic magnon branch, measured at 4 K with $E_i$ = 8 meV. A small anisotropy gap (~ 2 meV) is observed. All measured intensities are displayed in absolute cross section units (see Methods), and error bars indicate statistical uncertainty (one standard deviation).

Our highly extensive INS data allow us to determine the magnon spectrum over many Brillouin zones (BZs, Fig. 2c), in which we expect different dynamic structure factor $S(\mathbf{Q},\omega)$ but the same dispersion $\omega_m(\mathbf{q})$. Here, **Q** and ω are respectively the momentum and energy transfers of the scattering, *m* is the magnon branch index and **q** is the displacement of **Q** from the nearest BZ centre. Figure 3a-c displays raw INS data obtained along three different sets of high-symmetry lines in momentum space. The data are extremely clean, which gives us confidence in extracting $\omega_m(\mathbf{q})$ simply by inspection (Extended Data Fig. 3a). By fitting the experimental $\omega_m(\mathbf{q})$ using LSWT (see Methods and Extended Data Fig. 3), we conclude that interactions up to $J_9$ are necessary to describe the optical magnon dispersions. An effective anisotropy parameter is needed to describe the acoustic branch, but it does not affect the fitting of the optical branches. Using the optimized parameter set (Extended Data Table 1), we then calculate $S(\mathbf{Q},\omega)$ in all measured BZs (Fig. 3d-f), with the only remaining adjustable parameter being the size of the moments (0.78 $\mu_B/Cu^{2+}$, see Methods for detail). The agreement between the measurement and the calculation is astonishingly good, especially concerning the extremely rich structures in $S(\mathbf{Q},\omega)$. To our knowledge, this is probably by far the most successful LSWT description of experimentally observed magnons in a spin-1/2 system.

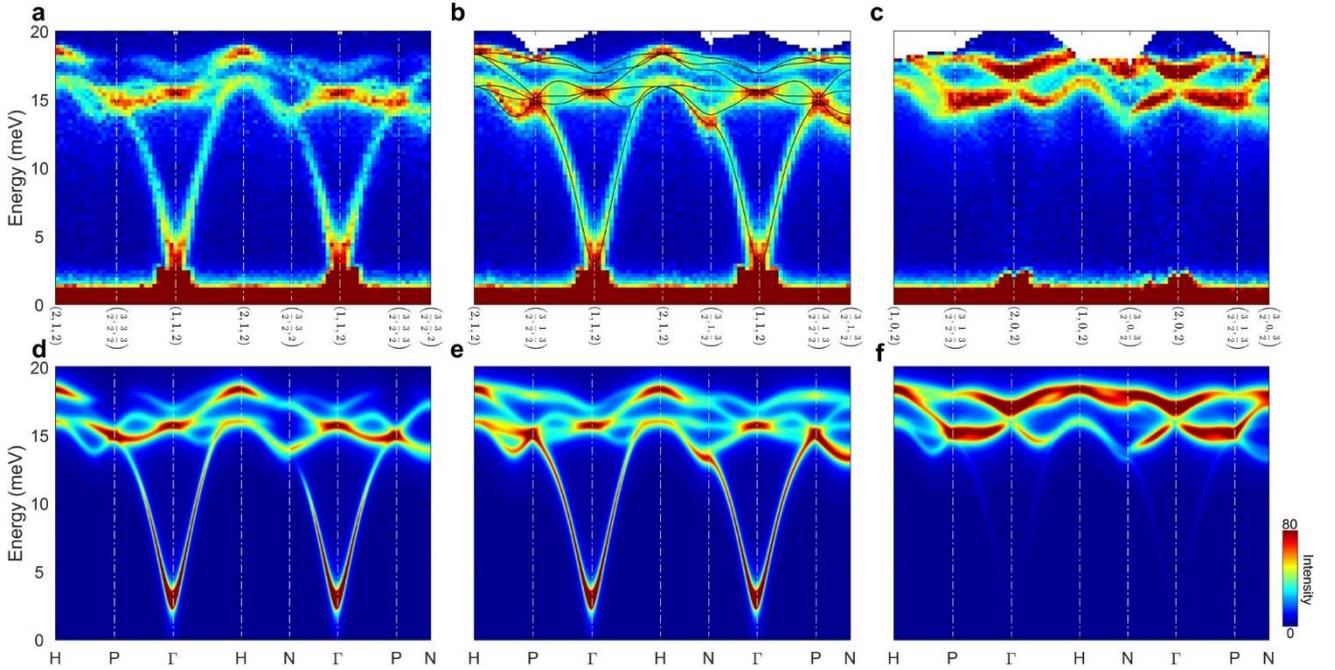

**Figure 3 | Comparison between measured and calculated magnon spectra. a-c,** INS intensities along three momentum trajectories indicated at the bottom, measured at 4 K with $E_i$ = 28 meV. Solid lines in **b** are calculated magnon dispersions using the parameter set that best describes the data (see Extended Data Fig. 3 and Extended Data Table 1). The same dispersions are applicable to the trajectories in **a** and **c**, which are equivalent in the reduced zone scheme. **d-f,** Calculated magnon $S(\mathbf{Q},\omega)$ along the same trajectories as in **a-c** using the same optimized parameter set. All measured INS intensities are displayed in absolute units (after Fig. 2), and all calculated intensities are displayed in the same units after dividing by a global rescaling factor of 1.28 (see Methods), which is used throughout the paper.

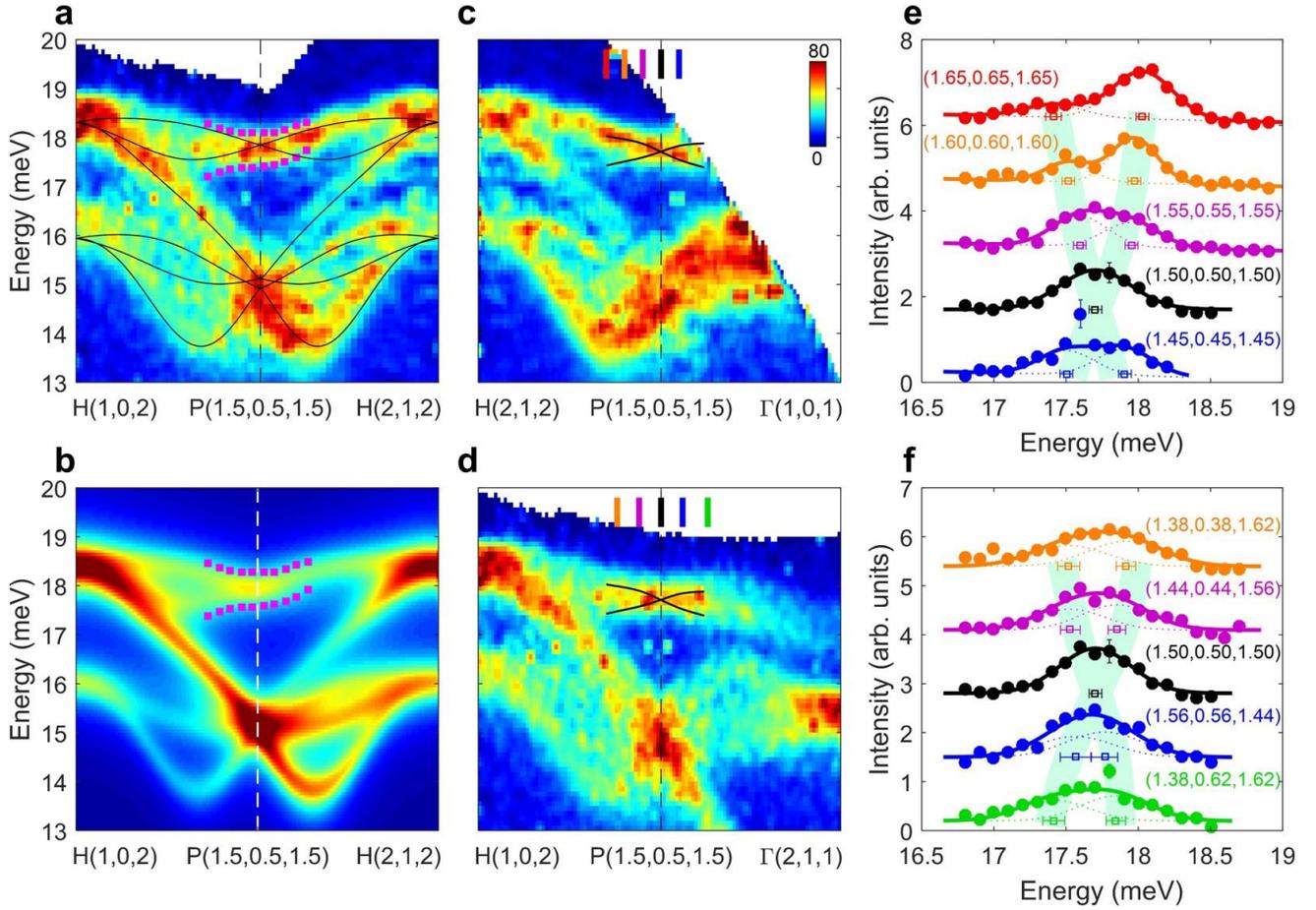

**Figure 4 | Evidence for Dirac-point-like magnon band crossing. a,** INS spectra along a H-P-H momentum trajectory. Solid lines are calculated magnon dispersions. **b,** Calculated magnon $S(\mathbf{Q},\omega)$ along the same trajectory. The magenta dotted lines (also in **a**) indicate an intensity envelope that is common between the calculation and the measurement. **c and d,** INS spectra along two different H-P-Γ momentum trajectories. **e and f,** EDCs (filled circles, color-coded with vertical bars in **c** and **d** indicating the **Q** positions) fitted with Gaussian peaks, offset for clarity. The fitted peak centres are indicated by open squares. Horizontal error bars indicate uncertainties in the peak positions, and vertical error bars indicate statistical uncertainty (one standard deviation; for most data points, it is comparable to the size of the symbols). Thick solid lines are guide to the eye to indicate the dispersion. They are the same in **e** and **f**, and are reproduced in **c** and **d** as black solid lines. All INS data in this figure were measured with $E_i$ = 31 meV at 4 K. The intensities in **a-d** are displayed in the same false-colour scale defined in the inset of **c** in absolute units (after Fig. 2).

We are now in a position to use the LSWT calculation to guide our search for topological band crossing in our INS data. According to the general theory [15], the P-point of the BZ (Fig. 2c) is always a crossing point (a Dirac point) when the system has U(1) spin-rotation symmetry. These Dirac points are indeed found in our calculated dispersions (Fig. 4a, solid lines): the six doubly-degenerate bands cross at three Dirac points located at the P-point at different energies. Two of them are very close together near 15 meV. The exact locations of these two Dirac points, as well as additional Dirac points nearby, sensitively depend on the details of the interactions (Extended Data Fig. 4) that fall beyond the precision of our measurements. Therefore, we focus here on the Dirac point between the topmost magnon bands at about 18 meV, which can be more clearly resolved.

We first compare INS spectra measured along momentum cuts that have the same $\omega_m(\mathbf{q})$ but different $S(\mathbf{Q},\omega)$. Data from different $\mathbf{Q}$ (but the same $\mathbf{q}$) highlight the bands in different fashions, and the contrast between them allows us to better identify the common dispersions. Figure 4a-b presents cuts connecting a P-point to two of its neighbouring H-points. In both the measurement and the calculation, the two bands near 18 meV are equally intense as $\mathbf{Q}$ moves away from P(1.5, 0.5, 1.5) towards H(1, 0, 2), whereas only the high-energy band is pronounced as $\mathbf{Q}$ moves towards H(2, 1, 2). This $\mathbf{Q}$ dependence of $S(\mathbf{Q},\omega)$ results in a distinct envelope shape of the magnetic intensity (magenta dotted lines in Fig. 4a-b). Similar comparisons when moving $\mathbf{Q}$ in other directions are presented in Extended Data Fig. 5. The contrast in the experimental $S(\mathbf{Q},\omega)$, together with the excellent agreement between the measurement and the calculation, demonstrates that there are two bands crossing at the P-point with a locally linear dispersion and no gap. The crossing is therefore consistent with being a three-dimensional Dirac point.

The presence of this Dirac-point-like crossing is further evinced by organizing the INS data into energy distribution curves (EDCs). Figure 4c&d displays data along a straight line in $\mathbf{Q}$ space, going from an H-point to a Γ-point via a P-point. Close to the P-point, we present EDCs at a series of $\mathbf{Q}$s (Fig. 4e&f). The EDCs are fitted by one or two Gaussian peaks, depending on whether the profile is consistent with a single peak that has a resolution-limited width (known to be about 0.58 meV full width at half maximum in these measurements). The fitted peaks disperse in a linear fashion through the P-point, with dispersion velocities consistent between the two different H-P-Γ cuts (Figs. 4c&e vs. d&f), and between the empirical fits (black lines in Fig. 4c&d) and the LSWT calculations (Fig. 4a). Based on this highly consistent set of evidence, we conclude that the topological magnon band crossing is confirmed in our experiment.

Even though a Dirac-point-like band crossing is fully consistent with our INS data, Dirac points are only the limiting case of nodal rings that carry $Z_2$ topological monopole charges [19], and they additionally require the presence of U(1) spin-rotation symmetry [15]. The U(1) symmetry in turn requires the magnetic ground state to be collinear, and the underlying spin Hamiltonian to be either isotropic (Heisenberg) or globally easy-axis *XXZ*. In the former case, the antiferromagnetic order breaks a continuous symmetry, so the spin excitations must be gapless, which is inconsistent with our experiment (Fig. 2e&f). The latter case is compatible neither with spin 1/2 nor with the high crystal symmetry of $Cu_3TeO_6$. Therefore, we believe that the U(1) symmetry must be broken in $Cu_3TeO_6$, which necessitates a slightly non-collinear magnetic ground state that is not inconsistent with neutron diffraction results [17]. Subsequently, the observed Dirac points must in fact be tiny nodal rings that are too small to be experimentally resolved. Demonstrating the possibility to have sizable nodal rings in other magnetic materials (with stronger U(1)-symmetry breaking interactions) will be interesting. In fact, magnon systems are superior to electron systems for finding such nodal rings. For electrons, the *PT* symmetry is required in conjunction with the absence of spin-orbit coupling, which is never strictly true, in order to protect such nodal rings [19]. Only the *PT* symmetry is required for magnons.

On top of this minimal requirement, additional symmetries, such as in the case of $Cu_3TeO_6$ here, may bring intriguing features to the magnon bands that deserve further investigation. The symmetry-enforced Dirac points (with U(1)) at the two P-points can be shown to have the same topological charges [15], so their presence necessitates the existence of additional Dirac points elsewhere in the BZ. Moreover, we discover a 'sum rule' of magnon energies at high symmetry points of the BZ: $\sum_m (\omega_{\Gamma,m}^2 + 4\omega_{P,m}^2 + \omega_{H,m}^2) = \sum_m 6\omega_{N,m}^2$, which imposes additional constraints on how the bands may cross into one another. The sum rule holds exactly in the LSWT (for models with at least nine *J*'s), and to a precision of about 1% in our measured dispersions. We believe that the sum rule is related to the space-group symmetry of the entire lattice, as well as to the site symmetry of $Cu^{2+}$. A close-knit comparison between real- and reciprocal-space pictures has led to a recent progress in the understanding of electronic band topology [29,30], where the high symmetry of $Cu_3TeO_6$ has been noted as an extreme case of interconnected bands [29]. A counterpart analysis for magnon states in the magnetic groups [20] may lead to new insights for the prediction of novel magnon systems.

**Acknowledgements**
We wish to thank Fa Wang, Gang Chen, Eugene Motoyama and Liang Fu for discussions and comments. The INS experiments were performed at the MLF, J-PARC, Japan, under a user program (proposal Nos. 2016B0116 and 2017I0001). The work at Peking University is supported by the National Natural Science Foundation of China (Grants No. 11374024 and No. 11522429) and Ministry of Science and Technology of China (Grants No. 2015CB921302 and No. 2013CB921903).

**Author Information**
Correspondence and requests for materials should be addressed to C.F. (cfang@iphy.ac.cn) and Y.L. (yuan.li@pku.edu.cn).

**References**

1. Bansil, A., Lin, H., & Das, T. Colloquium: Topological band theory. *Rev. Mod. Phys.* **88**, 021004 (2016).
2. Chiu, C.-K., Teo, J. C. Y., Schnyder, A. P., & Ryu, S. Classification of topological quantum matter with symmetries. *Rev. Mod. Phys.* **88**, 035005 (2016).
3. Burkov, A. A. Topological semimetals. *Nat. Mater.* **15,** 1145-1148 (2016).
4. Armitage, N. P., Mele, E. J., & Vishwanath, A. Weyl and Dirac semimetals in three dimensional solids. Preprint at https://arxiv.org/abs/1705.01111 (2017).
5. Murakami, S. Phase transition between the quantum spin Hall and insulator phases in 3D: emergence of a topological gapless phase. *New J. Phys.* **9,** 356 (2007).
6. Wan, X., Turner, A. M., Vishwanath, A., & Savrasov, S. Y. Topological semimetal and Fermi-arc surface states in the electronic structure of pyrochlore iridates. *Phys. Rev. B* **83,** 205101 (2011).
7. Young, S. M. *et al.* Dirac semimetal in three dimensions. *Phys. Rev. Lett.* **108,** 140405 (2012).
8. Liu, Z. K. *et al.* Discovery of a three-dimensional topological Dirac semimetal, $Na_3Bi$. *Science* **343,** 864-867 (2014).
9. Xu, S. Y. *et al.* Discovery of a Weyl fermion semimetal and topological Fermi arcs. *Science* **349,** 613-617 (2015).
10. Lv, B. Q. *et al.* Experimental discovery of Weyl semimetal TaAs. *Phys. Rev. X* **5,** 031013 (2015).
11. Lu, L., Joannopoulos, J. D., & Soljačić, M. Topological photonics. *Nat. Photonics* **8,** 821-829 (2014).
12. Stenull, O., Kane, C. L., & Lubensky, T. C. Topological phonons and Weyl lines in three dimensions. *Phys. Rev. Lett.* **117,** 068001 (2016).
13. Li, F. Y. *et al.* Weyl magnons in breathing pyrochlore antiferromagnets. *Nat. Commun.* **7,** 12691 (2016).
14. Mook, A., Henk, J., & Mertig, I. Tunable magnon Weyl points in ferromagnetic pyrochlores. *Phys. Rev. Lett.* **117**, 157204 (2016).
15. Li, K., Li, C., Hu, J., Li, Y., & Fang, C. Dirac and nodal-line magnons in collinear antiferromagnets. To appear in *PRL*. Preprint at https://arxiv.org/abs/1703.08545 (2017).
16. Lu, L. *et al.* Experimental observation of Weyl points. *Science* **349,** 622- 624 (2015).
17. Herak, M. *et al.* Novel spin lattice in $Cu_3TeO_6$: an antiferromagnetic order and domain dynamics. *J. Phys.: Condens. Matter* **17,** 7667-7679 (2005).
18. Månsson, M. *et al.* Magnetic order and transitions in the spin-web compound $Cu_3TeO_6$. *Physics Procedia* **30,** 142-145 (2012).
19. Fang, C., Chen, Y., Kee, H. Y., & Fu, L. Topological nodal line semimetals with and without spin-orbital coupling. *Phys. Rev. B* **92,** 081201(R) (2015).
20. Watanabe, H., Po, H. C., & Vishwanath, A. Structure and topology of band structures in

the 1651 magnetic space groups. Preprint at https://arxiv.org/abs/1707.01903 (2017).
21. Bradlyn, B. *et al.* Beyond Dirac and Weyl fermions: Unconventional quasiparticles in conventional crystals. *Science* **353**, aaf5037 (2016).
22. Watanabe, H., Po, H. C., Zaletel, M. P., & Vishwanath, A. Filling-enforced gaplessness in band strutures of the 230 space groups. *Phys. Rev. Lett.* **117**, 096404 (2016).
23. Bzdušek, T., Wu, Q., Rüegg, A., Sigrist, M., & Soluyanov, A. A. Nodal-chain metals. *Nature* **538**, 75-78 (2016).
24. Zhitomirsky, M. E., & Chernyshev, A. L. Colloquium: Spontaneous magnon decays. *Rev. Mod. Phys.* **85,** 219-243 (2013).
25. Lake, B., Tennant, D. A., Frost, C. D., & Nagler, S. E. Quantum criticality and universal scaling of a quantum antiferromagnet. *Nat. Mater.* **4,** 329-334 (2005).
26. Headings, N. S., Hayden, S. M., Coldea, R., & Perring, T. G. Anomalous high-energy spin excitations in the high-$T_c$ superconductor-parent antiferromagnet $La_2CuO_4$. *Phys. Rev. Lett.* **105,** 247001 (2010).
27. Dalla Piazza, B. *et al.* Fractional excitations in the square lattice quantum antiferromagnet. *Nat. Phys.* **11,** 62-68 (2015).
28. Whangbo, M. H., Koo, H. J., & Dai, D. Spin exchange interactions and magnetic structures of extended magnetic solids with localized spins: theoretical descriptions on formal, quantitative and qualitative levels. *J. Solid State Chem.* **176,** 417-481 (2003).
29. Bradlyn, B. *et al.* Topological quantum chemistry. *Nature* **547,** 298-305 (2017).
30. Po, H. C., Vishwanath, A., & Watanabe, H. Complete theory of symmetry-based indicators of band topology. *Nat. Commun.* **8,** 50 (2017).
31. He, Z., & Itoh, M. Magnetic behaviors of $Cu_3TeO_6$ with multiple spin lattices. *J. Mag. Mag. Mater.* **354,** 146-150 (2014).
32. Kajimoto, R. *et al.* The Fermi chopper spectrometer 4SEASONS at J-PARC. *J. Phys. Soc. Jpn.* **80,** SB025 (2011).
33. Nakamura, M. *et al.* First demonstration of novel method for inelastic neutron scattering measurement utilizing multiple incident energies. *J. Phys. Soc. Jpn.* **78,** 093002 (2009).
34. Inamura, Y., Nakatani, T., Suzuki, J., & Otomo, T. Development status of software "Utsusemi" for chopper spectrometers at MLF, J-PARC. *J. Phys. Soc. Jpn.* **82,** SA031 (2013).
35. Ewings, R. *et al.* HORACE: software for the analysis of data from single crystal spectroscopy experiments at time-of-flight neutron instruments. *Nucl. Instrum. Meth. A* **834,** 132-142 (2016).
36. Xu, G., Xu, Z., & Tranquada, J. M. Absolute cross-section normalization of magnetic neutron scattering data. *Rev. Sci. Instrum.* **84,** 083906 (2013).
37. Shirane, G., Shapiro, S. M., & Tranquada, J. M. Neutron scattering with a triple-axis spectrometer: Basic techniques. (Cambridge University Press, 2002).

38. Choi, K. Y., Lemmens, P., Choi, E. S., & Berger, H. Lattice anomalies and magnetic excitations of the spin web compound $Cu_3TeO_6$. *J. Phys.: Condens. Matter* **20,** 505214 (2008).
39. Herak, M. Cubic magnetic anisotropy of the antiferromagnetically ordered $Cu_3TeO_6$. *Solid State Commun.* **151,** 1588-1592 (2011).
40. Lorenzana, J., Seibold, G., & Coldea, R. Sum rules and missing spectral weight in magnetic neutron scattering in the cuprates. *Phys. Rev. B* **72,** 224511 (2005).
41. Sandvik, A. W. Finite-size scaling of the ground-state parameters of the two-dimensional Heisenberg model. *Phys. Rev. B* **56**, 11678-11690 (1997).
42. Schmidt, R., Schulenburg, J., Richter, J., & Betts, D. D. Spin-1/2 $J_1$-$J_2$ model on the body-centered cubic lattice. *Phys. Rev. B* **66**, 224406 (2002).

## Methods

### Sample growth and characterization

High-quality single crystals of $Cu_3TeO_6$ were grown by a flux method using molten $PbCl_2$ as solvent [31]. X-ray Laue backscattering from natural crystal surfaces produces sharp diffraction patterns with an approximate 4-fold symmetry (Extended Data Fig. 1a), consistent with the cubic space group $Ia$-3 (#206; a = 9.537 Å) [17]. For the INS experiments, we co-aligned about 80 pieces of single crystals by gluing them on aluminium plates using a hydrogen-free adhesive, amounting to a total crystal mass of about 16.8 grams (Extended Data Fig. 1a). The entire array has a total mosaic spread of about 2 degrees, as determined from the full widths at half maximum of rocking curves measured on the (0, 0, 3) and (2, 2, 0) Bragg reflections (Extended Data Fig. 1b). Temperature-dependent intensities of the magnetic Bragg reflection (1, 1, 0), as well as uniform magnetic susceptibility, indicate a sharp antiferromagnetic transition below $T_N$ = 61 K (Extended Data Fig. 1c-d). Fitting the high-temperature susceptibility data suggests a Curie-Weiss temperature of about -165 K, consistent with previous results [17].

### INS experiments

Our INS experiments were performed on the 4SEASONS time-of-flight spectrometer at the MLF, J-PARC, Japan [32]. The spectrometer has a multiple-$E_i$ capability [33], so that data in different energy ranges (with different energy resolutions) can be obtained simultaneously. All data presented were obtained with two chopper conditions: primary incident energy $E_i$ = 28 meV with chopper frequency 250 Hz (low resolution), and primary $E_i$ =31 meV with chopper frequency 400 Hz (high resolution). Two different sample orientations were used in our measurements, with crystallographic direction of either (1, 0, 0) or (1, 1, 0) being placed in the vertical direction. During the measurement, the sample is rotated about the vertical axis over a range of 180 degrees in steps of 0.5 degrees, and data accumulated at each angle were combined together, forming a four-dimensional data set, which we used the Utsusemi [34] and Horace [35] software packages to reduce and analyse. After a careful alignment of the measured data set with the crystallographic coordinate system using all available nuclear Bragg reflections, the entire data set was down-folded in the three-dimensional momentum space using the full cubic symmetry ($T_h$ point group, plus four-fold rotations about the <100> directions; the latter 'symmetries' were introduced by twinning during the crystal growth and the co-alignment processes). The folding resulted in a data volume that is 1/48 of the original, and it greatly enhanced the counting statistics by combining physically equivalent data points acquired by different detector pixels, without introducing any noticeable error. The recorded neutron intensities, first normalized by the amount of proton charge hitting the spallation target, were then compared against measurements of a vanadium standard sample using exactly the same spectrometer conditions, in order to convert the intensities to absolute scattering cross-section units [36]. The resultant cross sections were further corrected for neutron absorption, which is estimated to cause a minimum of 22% reduction of the scattering intensity based on tabulated data [37], $E_i$ = 28 meV, and an effective sample thickness of 18 mm. The absorption-corrected absolute cross sections are presented throughout the paper.

**LSWT fitting and simulations**

Although the collinear antiferromagnetic ground state of $Cu_3TeO_6$ can be readily understood by considering only the antiferromagnetic nearest-neighbour exchange interactions (Fig. 1a), spin interactions over longer distances turn out to be necessary for describing the observed spin excitations. We model the spin interactions as:

$$H = H_{1NN} + H_{2NN} + \cdots H_{MNN} = \sum_{d=1}^{M} J_d \sum_{\langle i,j \rangle \in dNN} \boldsymbol{S}_i \cdot \boldsymbol{S}_j,$$

where $J_d$ is the Heisenberg exchange interaction between the $d^{th}$-nearest neighbours. Once the number of interactions (*M*) and their strengths are chosen, standard Holstein-Primakoff transformation is performed, and the magnon dispersions can be obtained after a straightforward calculation [15].

Comparing the model to the measurement results, we can first rule out the *M* = 2 model. Under the notion that there are a total of six observed magnon branches (Figs. 2 and 3), the optical branches meet at a two-fold and a three-fold degenerate energy point at the Γ-point of the BZ, with the two-fold degenerate energy ($E_{\Gamma,2}$) higher than the three-fold degenerate energy ($E_{\Gamma,3}$). At the H-point, the six branches meet at two energies ($E_{H,+}$ and $E_{H,-}$), both of which are three-fold degenerate. Altogether, we have $E_{\Gamma,3} < E_{H,-} < E_{\Gamma,2} < E_{H,+}$, which turns out to be incompatible with the analytical expressions of the corresponding energies calculated from the *M* = 2 model:

$$E_{\Gamma,3} = \sqrt{4(-J_1 + J_2)(-J_1 + J_2)},$$

$$E_{\Gamma,2} = \sqrt{3(-J_1 + J_2)(-J_1 + 3J_2)},$$

$$E_{H,\pm} = \sqrt{3J_1^2 + 5J_2^2 - 8J_1J_2 \pm 4|J_2(J_1 - J_2)|}.$$

As *M* is further increased, analytical expressions at high-symmetry BZ points are no longer sufficient to determine the interactions. Therefore, we proceed by attempting to fit the magnon dispersions along high-symmetry momentum cuts. The high-quality INS data allow us to extract a discrete set of ω$_m$(**q**) points along the high-symmetry lines, as displayed in Extended Data Fig. 3a. As our main goal here is to use LSWT calculations to guide our search for topological magnon band crossing, we have purposely refrained from introducing band-crossing structures into the extracted ω$_m$(**q**) data, in order to avoid biasing the model.

We then perform nonlinear least-squares fitting of the ω$_m$(**q**) data by the Levenberg-Marquardt method. To overcome local-minima problems in the fitting process, we have performed a systematic search by starting from a multi-dimensional grid of the initial parameter set, and used $\chi^2$ to assess the goodness of the fit obtained from each initial parameter set before a globally optimized result is obtained. The experimental dispersions cannot be well-described by the $M = 6$ model (Extended Data Fig. 3b), but the quality of the fit is much improved with $M = 7$ (Extended Data Fig. 3c), which results in a parameter set that is dominated by $J_1$ and $J_7$. However, structural considerations indicate that the exchange pathway of $J_9$ is even more favourable for a strong interaction than that of $J_7$, see Fig. 1b and Extended Data Fig. 2, as well as Extended Data Table 1. Therefore, we have further extended the model to $M = 9$. Indeed, not only do we find that the $M = 9$ model is more likely to converge to parameters dominated by $J_1$ and $J_9$, but the fit quality is noticeably improved (Extended Data Fig. 3d&e). Thus we conclude that the $M = 9$ model dominated by $J_1$ and $J_9$ is the most suitable description of the spin interactions in $Cu_3TeO_6$. This result is very different from all previous understandings of the spin interaction network of this compound [17,18,31,38].

Although the $M = 9$ Heisenberg model successfully describes the optical magnon dispersions, a noticeable discrepancy from the experimental data is the lack of low-energy excitation gap at the BZ centre. This is expected because the antiferromagnetic order breaks the continuous SU(2) symmetry, which guarantees that the low-energy excitations are gapless Goldstone modes. A physically rigorous remedy to this discrepancy is to introduce site-dependent exchange anisotropy that respects the crystal symmetry, such as Dzyaloshinsky-Moriya interactions [15,39], because spin-1/2 systems cannot have single-ion anisotropy. However, the presence of such site-dependent exchange anisotropy generally favours a non-collinear magnetic structure, which significantly complicates the LSWT calculations. Meanwhile, neutron powder diffraction results indicate that the magnetic order in $Cu_3TeO_6$ is predominantly collinear, with possible non-collinear canting of the spins being no more than 6 degrees [17]; moreover, our INS data suggest that all magnons are two-fold degenerate (6 instead of 12 branches), which indicates that the magnetic ground state is approximately collinear. Therefore, we believe that the experimentally observed anisotropy gap can be accounted for by introducing a phenomenological global single-ion anisotropy term $H_a = -D(S_i^z)^2$ ($D > 0$), without affecting the description of the optical branches. For the $M = 9$ model, this term leads to a gap at the Γ-point, $\Delta = \frac{1}{2}\sqrt{D(8J_1 + 4J_3 + 4J_5 + 8J_7 + 8J_9 + D)}$.

Indeed, Extended Data Fig. 3f shows that the successful description of the optical magnon dispersions remains intact after the anisotropy gap has been accounted for, and the best-fit parameters after introducing this anisotropy are very similar to those in the Heisenberg model (Extended Data Table 1). We note that such global single-ion anisotropy does not break the

U(1) symmetry, hence our anisotropic model still results in Dirac-point-like band crossings rather than nodal rings, which are generally expected with the more realistic site-dependent exchange anisotropy [15]. However, given the very little effect of the single-ion anisotropy term on the optical magnon dispersions, we believe that the exchange anisotropy in $Cu_3TeO_6$ would not lead to any observable consequences in the optical magnon dispersions either.

Finally, to compare with the experimental data, we calculate the excitation spectra at any general **Q** and ω using the fully-optimized parameter set of the $M = 9$ model with global single-ion anisotropy, by calculating

$$S^{\alpha\beta}(\mathbf{Q},\omega) = \frac{1}{2\pi}\int_{-\infty}^{\infty} dt\, e^{-i\omega t}\sum_l e^{i\mathbf{Q}\cdot\mathbf{r}_l}\langle S_0^\alpha(0)S_l^\beta(t)\rangle,$$

which can be converted into absolute scattering cross section units:

$$\frac{k}{k'}\frac{d^2\sigma}{d\Omega dE} = \frac{N}{\hbar}\left(\frac{\gamma r_e}{2}\right)^2 g^2|F(\mathbf{Q})|^2 e^{-2W(Q)}\sum_{\alpha\beta}(\delta_{\alpha\beta}-Q_\alpha Q_\beta)S^{\alpha\beta}(\mathbf{Q},\omega),$$

where $N$ is the number of primitive cells in the sample, $\left(\frac{\gamma r_e}{2}\right)^2 = 72.65\times 10^{-3}$ barn, $g$ (= 2) is the Landé splitting factor, and $\alpha$ and $\beta$ are indices (*xyz*) of a Cartesian coordinate system [40]. For simplicity, we assume the Debye-Waller factor $e^{-2W(Q)}$ to be unity, and calculate the magnetic form factor $F(\mathbf{Q})$ in the isotropic approximation (for our measured momentum region, $|F(Q)|^2$ amounts to about 0.75). To reproduce the measured INS spectra, we perform the same **Q**-space folding of the calculated $S(\mathbf{Q},\omega)$ and use $\langle 1-Q_\alpha^2\rangle_{domain} = \frac{2}{3}$, both of which account for the presence of multiple antiferromagnetic domains in our sample [17, 39] and the fact that the neutron beam is not spin polarized. A damping to the magnons proportional to the magnon energies has been introduced so that the calculated magnon intensities have a finite energy width rather than being delta-function-like singularities. The calculated absolute cross sections are globally greater than the measured ones by a factor of 1.28, which effectively means that the measured intensities correspond to a spin system with $S = 0.39$ instead of $S = 1/2$. Since we have not considered the Debye-Waller factor, and because we have only corrected the measured intensities for neutron absorptions using the incident energy (the scattered neutrons are less energetic and suffer more from absorption), the above estimate of $S = 0.39$ constitutes a lower bound of the size of moment on each $Cu^{2+}$ (0.78 μB/$Cu^{2+}$). Note that this moment size is considerably larger than the expected value (0.60 μB) in a two-dimensional square lattice [41], and is very close to that (0.83 μB) in a body-centred cubic lattice [42], which has the same coordination number ($N = 8$) as the magnetic lattice of $Cu_3TeO_6$. The presented calculated intensities have been adjusted to the case of $S = 0.39$ throughout the manuscript. After applying this global factor for the intensity, the agreement between the calculated and the measured spectra is remarkable, especially given the complexity of $S(\mathbf{Q},\omega)$.

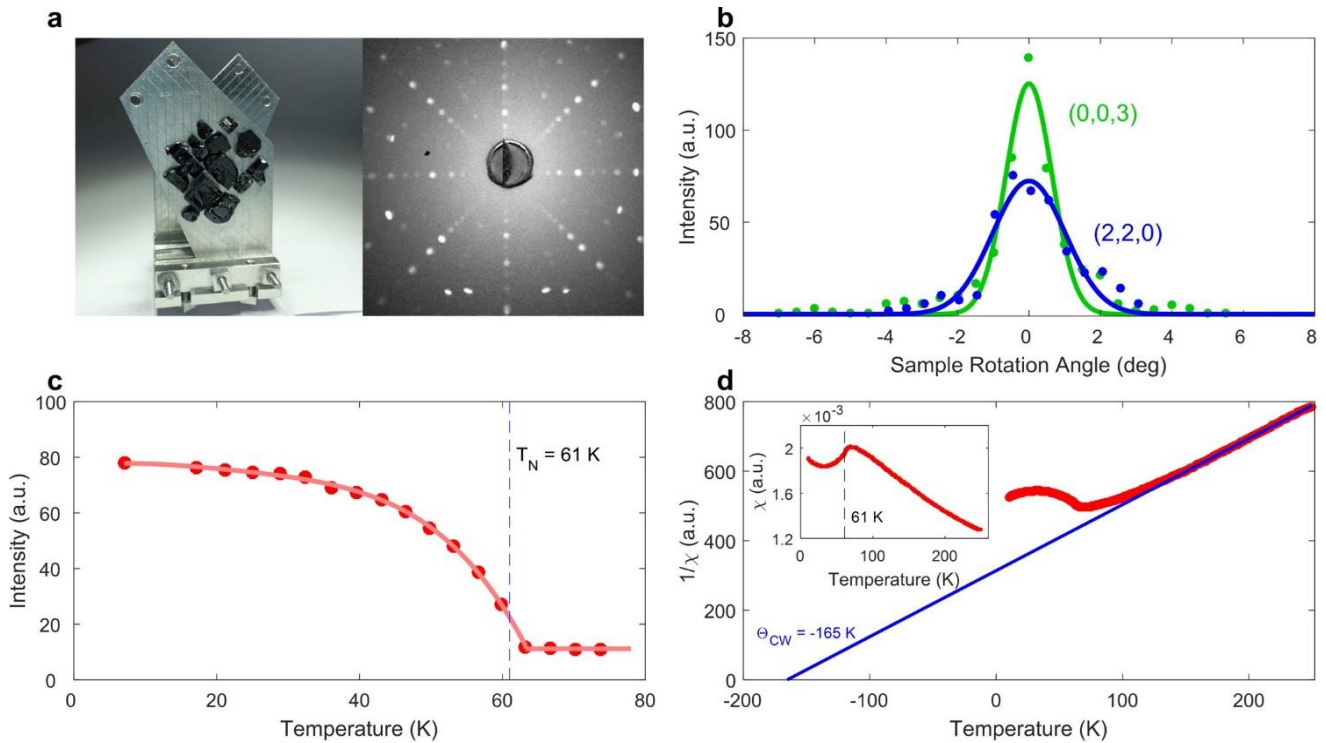

**Extended Data Figure 1 | Sample preparation and characterization. a,** Left side: photograph of $Cu_3TeO_6$ single crystals co-aligned on an aluminium sample holder. Right side: representative X-ray Laue pattern taken on a natural surface of a single crystal. **b,** Neutron diffraction intensities of selected Bragg reflections recorded upon rotating the entire sample array. Solid lines are Gaussian fits to the data, with full width at half maximum equal to 2.4 and 1.5 degrees for the (2, 2, 0) and (0, 0, 3) reflections, respectively. **c,** Magnetic neutron diffraction intensity measured on the (1, 1, 0) reflection as a function of temperature, indicating a clear antiferromagnetic transition below $T_N$ = 61 K. **d,** Uniform magnetic susceptibility measured on a single crystal using a Quantum Design MPMS. The data indicate an antiferromagnetic transition below 61 K and a Curie-Weiss temperature of about -165 K, consistent with previous reports [17, 31].

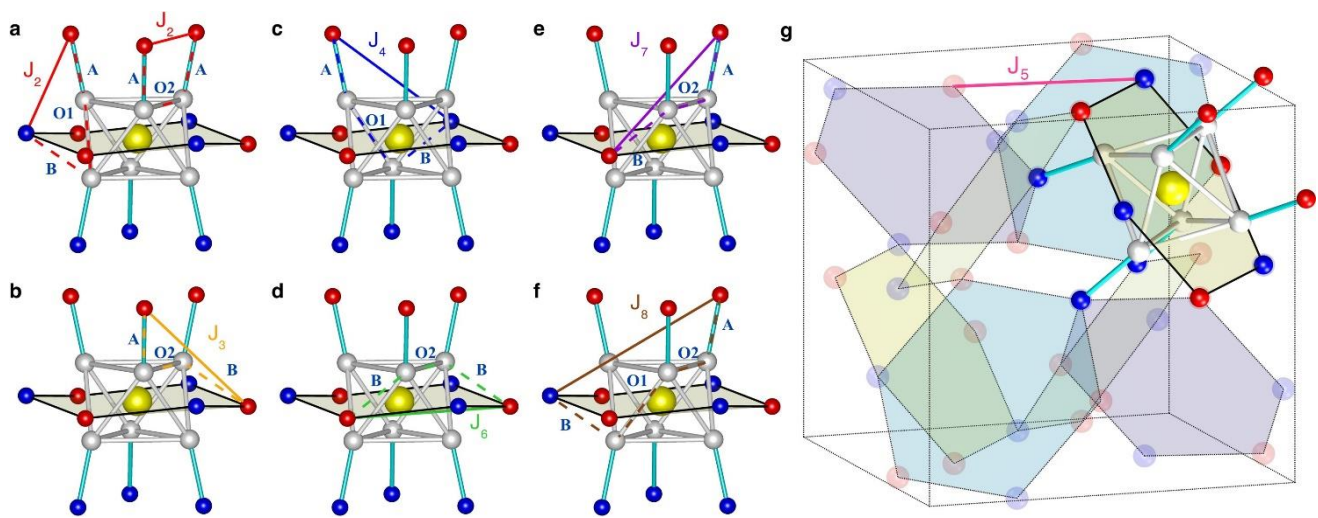

**Extended Data Figure 2 | Exchange pathways of interactions from $J_2$ to $J_8$. a-f,** The exchange pathways are indicated by dashed lines and labelled in accordance with Extended Data Table 1. $J_2$ involves two types of pathways, which may separately give rise to antiferromagnetic and ferromagnetic interactions of comparable strengths. A cancellation between these two contributions may explain the total weakness of $J_2$. **g,** $J_5$ is displayed directly in the unit cell. It involves a different type of O-O bond which cannot be accommodated into the structural unit shown in **a-f**.

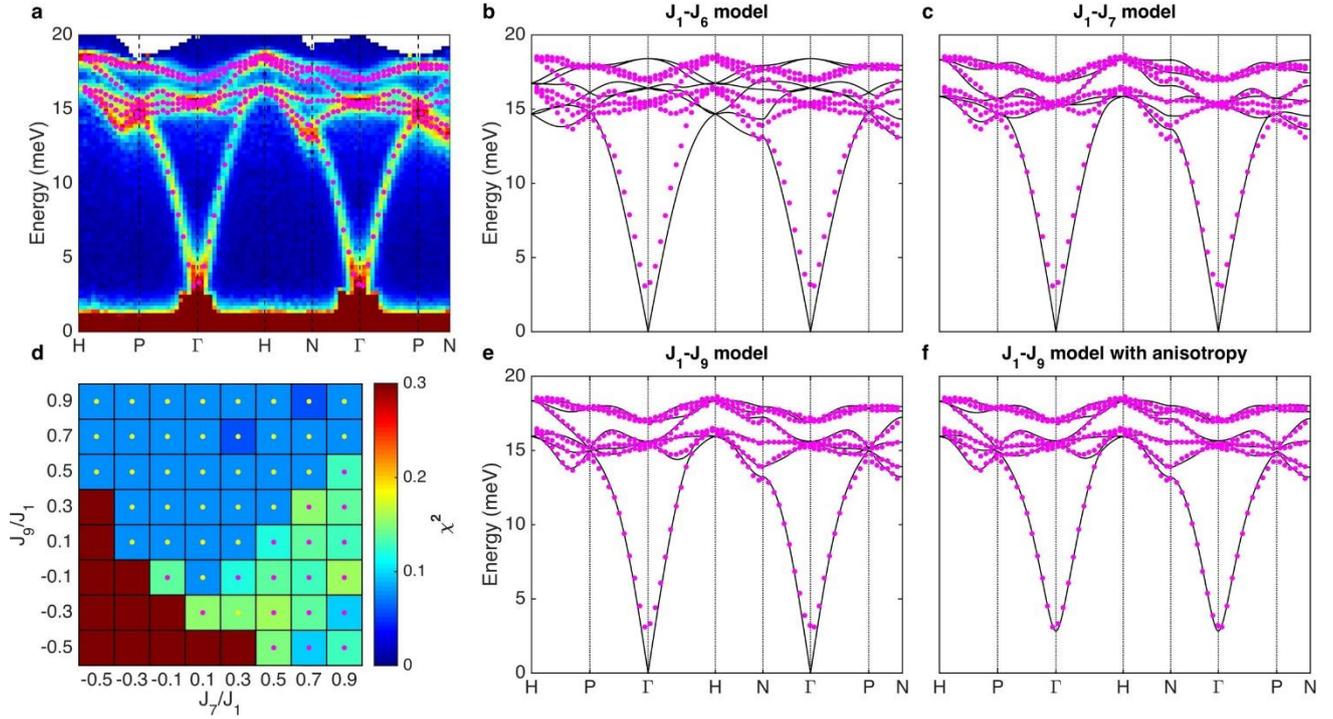

**Extended Data Figure 3 | Fitting magnon dispersions with LSWT. a,** Discrete points of $\omega_m(\mathbf{q})$ extracted by inspecting the experimental spectra. Although shown along only one momentum trajectory (the same as the one in Fig. 3b), data from other trajectories have been considered collectively in a conservative fashion, *i.e.*, we have purposely refrained from introducing band-crossing structures into the readings of $\omega_m(\mathbf{q})$. **b,** Best-fit result with interactions up to $J_6$, which clearly fails to reproduce the experimental data. **c,** Best-fit result with interactions up to $J_7$, which is substantially improved from the result in **b**. **d,** Comparison of fit results with interactions up to $J_9$, starting from various $J_7$-$J_9$ combinations. Reasonably good fits are obtained when at least one of them starts off large, so that the fitting converges to primarily "$J_1$-$J_7$" (magenta dots) or "$J_1$-$J_9$" (yellow dots) models. The latter model fits the data better, as indicated by the $\chi^2$ values displayed in colour. **e,** Best-fit result with interactions up to $J_9$. **f,** Best-fit result with interactions up to $J_9$ and an effective global single-ion anisotropy (see Methods).

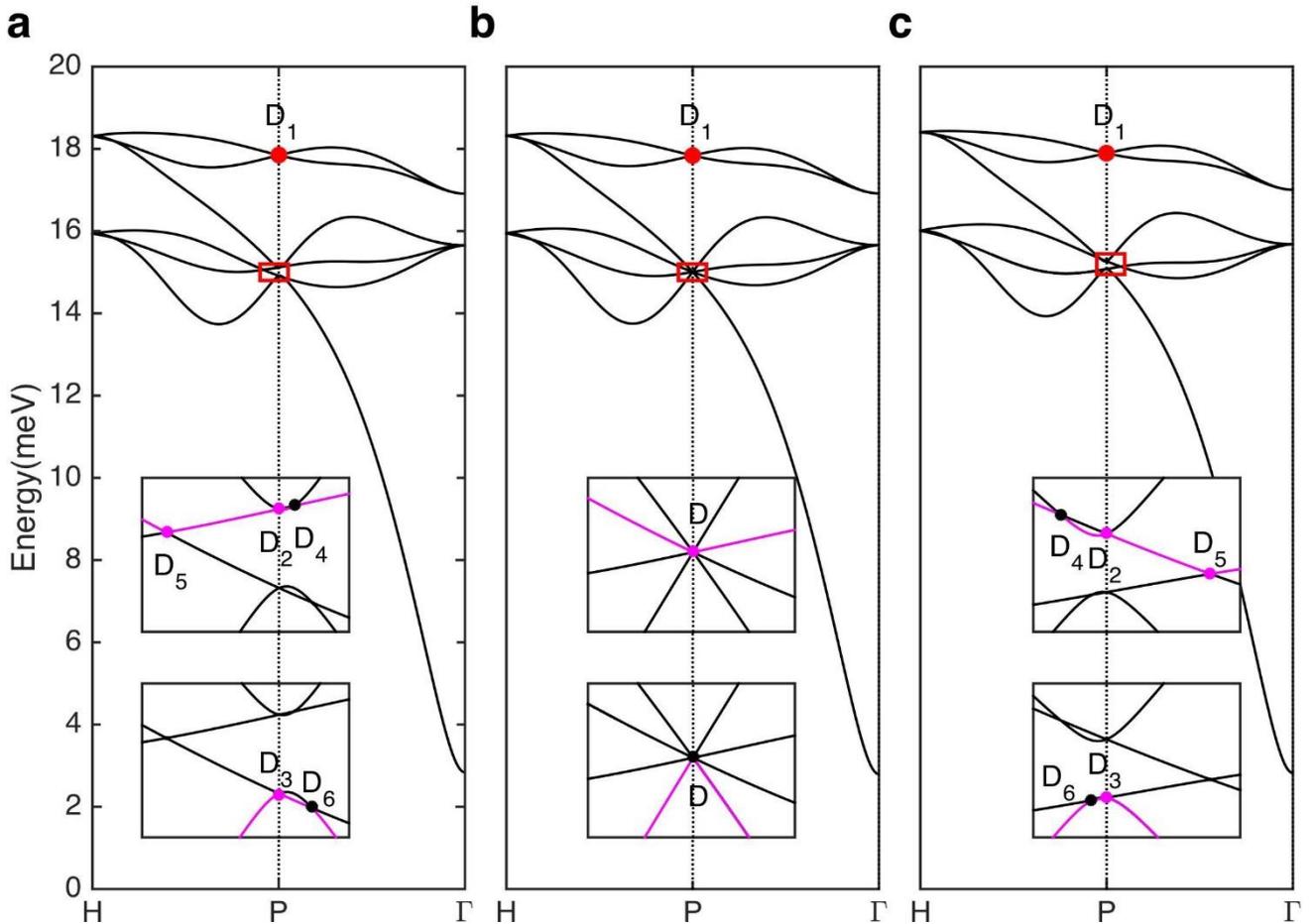

**Extended Data Figure 4 | Dirac points at and near the P-point.** $D_1$ is always a Dirac point regardless of model details, whereas the crossings near 15 meV depend on the sign of $J_2-J_6+J_8$. The insets display zoom-in views near 15 meV, where magenta and black circles indicate positive and negative monopole charges, respectively, of the Dirac points [15] for the band highlighted in magenta. **a,** $J_2-J_6+J_8 = -0.12$ meV. Topological charges: $D_2= +1$, $D_3= +1$, $D_4= -1$ (x4), $D_5 = +1$ (x4), $D_6= -1$ (x4), where the numbers in the parentheses indicate the number of symmetry equivalents around the given P-point. **b,** $J_2-J_6+J_8=0$. We denote the four-band touching point as D. In the upper inset, the topological charge at D is +1, which equals to $D_2+4(D_4+D_5)$ after considering all symmetry equivalents. In the lower inset, the topological charge at D is -3, which equals to $D_3+4D_6$. **c,** $J_2-J_6+J_8 = 0.11$ meV. The topological charges are the same as those in **a**, although the Dirac points appear at different momentum positions.

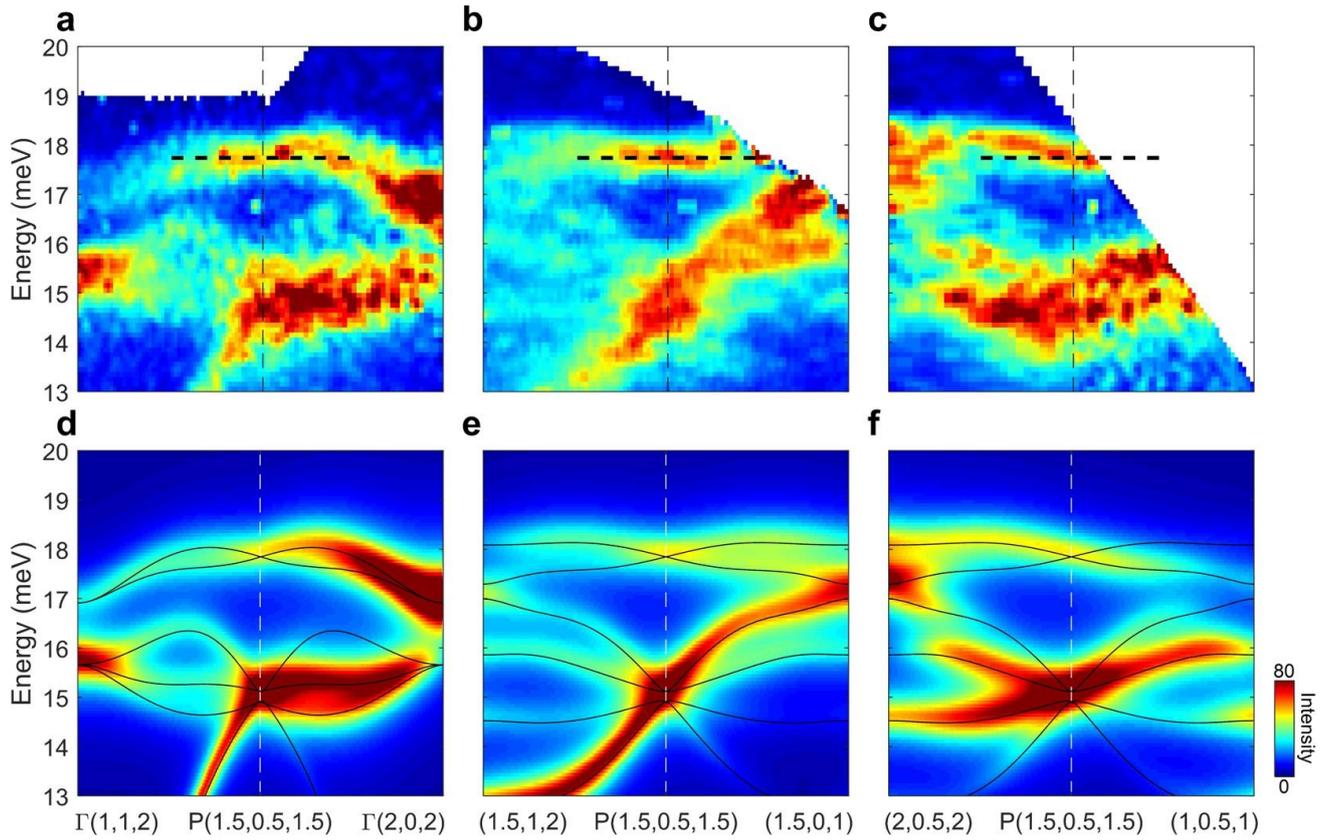

**Extended Data Figure 5 | Optical magnon dispersions near the P-point. a-c,** INS data obtained along three different momentum cuts through the same P-point. The intensities are displayed in absolute units after Fig. 2. **d-f,** LSWT calculation results for the same momentum cuts. Dashed lines in **a-c** are reference to the eye, in order to contrast the differently highlighted bands between the left and the right sides of **a**, and between the left sides of **b** and **c**. These data, even without comparison to the LSWT calculations, indicate the presence of two magnon bands crossing at the P-point without opening a gap. Solid lines in **d-f** are calculated magnon dispersions.

| Interaction | | Distance (nm) | Bond sequence and angles | Strength (meV) (without anisotropy) | Strength (meV) (with effective anisotropy) |
|---|---|---|---|---|---|
| $J_1$ | | 0.318 | A ∠106.2° B | **4.49** | **4.39** |
| $J_2$ | $J_{2A}$ | 0.360 | A ∠145.8° O1 ∠59.7° B | -0.22 | -0.36 |
| | $J_{2B}$ | | A ∠102.1° O2 ∠101.2° A | | |
| $J_3$ | | 0.477 | A ∠101.2° O2 ∠100.8° B (x2) | -1.49 | -1.61 |
| $J_4$ | | 0.481 | A ∠147.5° O1 ∠102.6° B (x2) | 1.33 | 1.30 |
| $J_5$ | | 0.481 | not available | 1.79 | 1.47 |
| $J_6$ | | 0.548 | B ∠100.8° O2 ∠149.5° B | -0.21 | -0.21 |
| $J_7$ | | 0.573 | A ∠102.1° O2 ∠149.5° B | -0.14 | -0.20 |
| $J_8$ | | 0.597 | A ∠102.1° O2 ∠90.0° O1 ∠102.6° B (x2) | 0.11 | 0.03 |
| $J_9$ | | 0.621 | A ∠145.8° O1 ∠147.5° A | **4.51** | **4.51** |

A: shorter Cu-O bond: 0.195 nm  B: longer Cu-O bond: 0.203 nm
O1: shorter O-O bond: 0.262 nm  O2: longer O-O bond: 0.281 nm

**Extended Data Table 1 | Detailed information about the exchange interactions.** Structural properties including Cu-Cu distance, chemical bonds and bond angles are presented in the second and third columns. The exchange pathway of $J_5$ cannot be described in a similar fashion, because it involves yet another type of O-O bond (length = 0.271 nm) that passes by additional $Cu^{2+}$. For some of the interactions, pathways involving a third type of Cu-O bond (length = 0.237 nm) have not been included in the table because of the expected weakness of the bond. LSWT fitting results (see text and Methods) obtained without and with global single-ion anisotropy (fitted to be $D = 0.46$ meV with anisotropy gap = 2.8 meV; this is unphysical for spin 1/2, but is nevertheless used as an effective parameter) are displayed in the last two columns. In both cases, $J_9$ and $J_1$ are determined to be the primary interactions. The fit results are overall consistent with the understanding that fewer sequential bonds and straighter bond angles lead to stronger antiferromagnetic interactions, and that bond angles close to 90 degrees favour ferromagnetic interactions. $J_2$ is particularly weak despite the short Cu-Cu distance, probably because the two associated exchange pathways ($J_{2A}$ and $J_{2B}$) cancel against each other.